\documentclass[onecolumn]{IEEEtran}
\IEEEoverridecommandlockouts
\usepackage{cite}
\usepackage{amsmath,amssymb,amsfonts}
\usepackage{algorithmic}
\usepackage{graphicx}
\usepackage{textcomp}
\usepackage{epstopdf}
\usepackage{threeparttable,booktabs}
\usepackage{xcolor}
\usepackage{url}
\begin{document}

\title{Improving Networked Music Performance Systems Using Application-Network Collaboration}

\author{\IEEEauthorblockN{\textbf{Emmanouil Lakiotakis}}
\IEEEauthorblockA{\textit{} \\
\textit{Foundation for Research and Technology Hellas}\\
aaa@sss.com}\\ 
\and
\IEEEauthorblockN{\textbf{Christos Liaskos}}
\IEEEauthorblockA{\textit{} \\
\textit{Foundation for Research and Technology Hellas}\\
cliaskos@ics.forth.gr}\\
\and
\IEEEauthorblockN{\textbf{Xenofontas Dimitripoulos}}
\IEEEauthorblockA{\textit{} \\
\textit{Foundation for Research and Technology Hellas and University of Crete}\\
fontas@ics.forth.gr} \\
}

\maketitle

\begin{abstract}
Networked Music Performance (NMP) systems involve musicians located in different places who perform music while staying synchronized via the Internet. The maximum end-to-end delay in NMP applications is called Ensemble Performance Threshold (EPT) and should be less than 25 milliseconds. Due to this constraint, NMPs require ultra-low delay solutions for audio coding, network transmission, relaying and decoding, each one a challenging task on its own. There are two directions for study in the related work referring to the NMP systems. From the audio perspective, researchers experiment on low-delay encoders and transmission patterns, aiming to reduce the processing delay of the audio transmission, but they ignore the network performance. On the other hand, network-oriented researchers try to reduce the network delay, which contributes to reduced end-to-end delay. In our proposed approach, we introduce an integration of dynamic audio and network configuration to satisfy the EPT constraint. The basic idea is that the major components participating in an NMP system the application and the network interact during the live music performance. As the network delay increases, the network tries to equalize it by modifying the routing behavior using Software Defined Networking principles. If the network delay exceeds a maximum affordable threshold, the network reacts by informing the application to change the audio processing pattern to overcome the delay increase, resulting in below EPT end-to-end delay. A full prototype of the proposed system was implemented and extensively evaluated in an emulated environment.
\end{abstract}


\section{Introduction}\label{sec:introduction}

A large number of daily life applications require Internet connectivity. These services include instant messaging, web browsing, multimedia communications, financial transactions etc. In all cases, the Internet is used as the medium for communication and data exchange between users. Based on the nature and the requirements each application has, they can be grouped into subcategories. For instance, multimedia streaming services require Internet connection without packet losses and minimized network delay due to congestion. On the other hand, file transfer services focus on faultless transmission but they afford network delay. In this work, we introduce the interaction between an application and the network, emphasizing on Networked Music Performance (NMP) systems.

A class of services that attracts the interest of the research community is real time applications. Users are usually located in different places around the world and their interaction should not be affected by distance as they should behave as being in the same room. The major indicator for performance in real time applications is end-to-end delay, that should be minimized in case of ultra-low delay sensitive applications. In this study we focus on a specific class of real-time applications called Networked Music Performance (NMP) systems. This term was firstly initiated by John Lazzaro from Berkeley University in 2001 and since then this term describes the distant music interaction in real time via the Internet \cite{carot2007network}.

NMP systems differ from the other multimedia real time applications. Due to its nature, the musicians that perform via the NMP should interact instantly, showing minimal delay tolerance \cite{xylomenos2013reduced}. In more detail, the maximum affordable delay between the initially played and the finally received signal should be less or equal to 25 milliseconds. This constraint is named as Ensemble Performance Threshold (EPT) among the audio-oriented researchers \cite{chafe2004effect}. As a result, NMP systems constitute a challenging research topic in the ultra-low delay sensitive applications and each parameter that affects end-to-end delay should be examined in order to achieve required performance.

In this work, a detailed study of NMPs takes place, containing the basic factors that play key role in NMP performance and a proposed approach that overcomes problems that take place during the live performance. The basic idea is that our approach introduces the collaboration between the application side and the network in order to deal with the network congestion problem. In more detail, our architecture exploits the flexibility that Software Defined Networking offers in changing network policies dynamically in order to choose the path with the lowest network delay among the available paths between the transmitter and the receiver. In case that all available paths are congested, the network informs the application side to change fundamental audio parameters such as sampling rate and frame size in order to decrease the delay created by audio processing. This change is crucial since the end-to-end delay in the NMP systems is the summary of the network delay and the audio processing delay. For this reason, choosing an audio mode with lower processing delay can afford further network delay increase, resulting in affordable end-to-end delay for the NMP system.

The rest of the paper is organized as follows. In section \ref{sec:background} we provide all the background knowledge about the  NMPs. In section \ref{sec:system_design} we describe the proposed architecture and examine remarkable use cases. In section \ref{sec:evaluation}, we present the emulation-based results produced during the performance evaluation of our approach. In section \ref{sec:related_work}, a detailed description about the state of the art in NMP area takes place containing also similar research works. Finally, in section \ref{sec:conclusions}, we emphasize on the conclusions from our approach, as well as possible future extensions.

\section{Prerequisites}
\label{sec:background}
The current section contains an extended analysis of the NMP services, the fundamental components used in conventional NMPs and problems during the NMP operation.

NMP services target to allow musicians located in different geographic locations performing together as described in section \ref{sec:introduction}. Capturing, processing and transmission stages during the NMP event can introduce significant packet and processing delays, that can cause problems for the synchronicity during the music performance. Due to the wide development of the Internet in recent decades, new opportunities arise for real time music interaction. In \cite{barbosa2003displaced}, researchers introduce a categorization about computer-assisted music services. The first class of these services describes local music networks, that are interconnected with each other and allow simultaneous interaction between the musicians via virtual instruments \cite{blaine2002jam}. The next type describes music systems used for asynchronous exchange, editing MIDI (Musical Instrument Digital Interface) audio data and building virtual rooms for audio recording and transmission using centralized servers \cite{latta1991notes,vlachakis2014using,akoumianakis2013collaborative}. The third class contains sonic environments with many music instrument players participating in improvisation experiments as well as audience during the process \cite{duckworth1999making,mckinney2012yig,barbosa2005network}. The last category describes real time musical interaction between musicians located in different places, requiring perfect synchronicity, resulting in Networked Music Performance systems. Some applications that belong to NMPs are tele-auditions, remote music teaching and distributed jam sessions and concerts \cite{rottondi2016overview}.

NMPs try to replace the physical presence of musicians in the same room with musicians interconnected using the Internet. In conventional live concerts, where the musicians perform at same location, they use visual feedback as well as gestures for keeping synchronized with the each other \cite{woszczyk2005shake}. Due to the strict requirements in terms of latency and jitter that NMPs have, these factors cannot be replaced in remote musical performances.

In many approaches \cite{carot2009fundamentals}, in conventional concerts, the maximum affordable delay is evaluated between 20 and 30 milliseconds, given a distance about 9 meters between musicians and considering the speed of sound propagation in the air. This delay tolerance threshold is crucial for live music concerts in order to keep synchronicity with the conductor. In fact, this threshold depends on the ability of the musician, his experience level and the used technique in order to combine the  self-delayed sound and the sound of the other musicians. Also, this threshold relies on the nature of the instrument. Different type of music instruments have different synchronicity requirements. In the research community, delay values between 30 to 50 milliseconds are accepted, depending on the distance between the musicians, the environment of the concert, the number and the type of the music instruments that participate in the concert and the musician experience \cite{lago2004quest,maki2005musical}. For instance, organ players should take into account the delay between pressing the keys and hearing the created sound, considering the physical distance between the keyboards and the pipes of the instrument. Similarly, a piano player should cope with the delay between 30 and 100 milliseconds, due to the piano audio attributes \cite{askenfelt1990touch}.

The factors described previously are met in a concert where the musicians are physically present. In Networked Music Performance systems, the maximum affordable delay is defined by the term called Ensemble Performance Threshold (EPT) among the audio-oriented researchers \cite{chafe2004effect}. EPT indicates that the maximum affordable delay between the initially played and the finally received signal should be less or equal to 25 milliseconds. The main objective in NMPs is the required synchronicity between multiple audio sources. The synchronicity depends on many factors. First of all, an NMP combines audio streams for different musicians that use computers with different clocks. The difference between clocks can result in under-run (when the sound-card at the receiver is fed with audio date with lower bit-rate than that applied when the application receives data from the network, which makes the application to wait until the buffer is refilled) or over-run phenomena (when the buffer receives data with higher bit-rate than that the application uses to process the data and this causes loss for the incoming data). For this reason, advanced signal processing approaches (e.g truncating or padding audio data) are applied in order avoid quality problems. Similar approaches are used in case of poor network performance that causes increased delay and jitter or packet loss.

NMP systems consider two possible causes that result in end-to-end delay: the first cause refers to the delay created due to audio processing. This refers to the audio capturing process from the audio hardware (sound-cards), the audio encoding/decoding processes in the transmitter/receiver side and the audio signal fragmentation into packets before transmission starts. The second cause is related to the delay created by the network during the audio transmission between the transmitter and  the receiver. The transmitted packets pass from hundreds of routers and switches towards the final destination, based on the routing policies. In each intermediate node, the packets are stored temporarily in queues and then they are forwarded after a period of time based on the workload due to the inbound traffic for each node. This results in the jitter phenomenon during the transmission process and explains why the network delay is higher than the real propagation delay caused by the physical distance between peers \cite{cart2006network}. Also, the available bandwidth plays an important role in the end-to-end delay since conventional Internet connections like DSL offer network delay higher than EPT constraint, which prevents the deployment of NMP systems.

From the audio perspective, due to the restricted bandwidth availability, researchers incorporate audio encoding approaches in order to reduce bit-rate in required levels but conventional audio coders increase the latency due to the heavy computational complexity that encoding/decoding processes insert. For instance, well known MP3 and AAC encoders create at least 100 ms delay \cite{goto1910virtual}. As a result, NMPs avoid using encoding/decoding techniques due to the increased latency. A solution that offers encoding/decoding capability with smaller delay comes from ultra-low delay coders that allow for modifying audio signal parameters such as the frame size. The most well known approaches in this category are the Advanced Audio Coding-Low Delay (AAC-LD) algorithm \cite{lutzky2005structural}, the Ultra Low Delay (ULD) audio coding algorithm \cite{wabnik2009error}  and the Constrained Energy Lapped Transform (CELT) codec, recently merged into the Opus Codec \cite{valin2016high}. All these approaches offer ultra-low delay coding and adjustable bit-rate levels. On the other hand, even ultra-low coders offer delay around 20 ms which is usually prohibitive for the NMPs since they insert a significant amount of the affordable end-to-end delay.

\begin{figure}[t]
\centerline{\includegraphics[width=0.58 \textwidth]{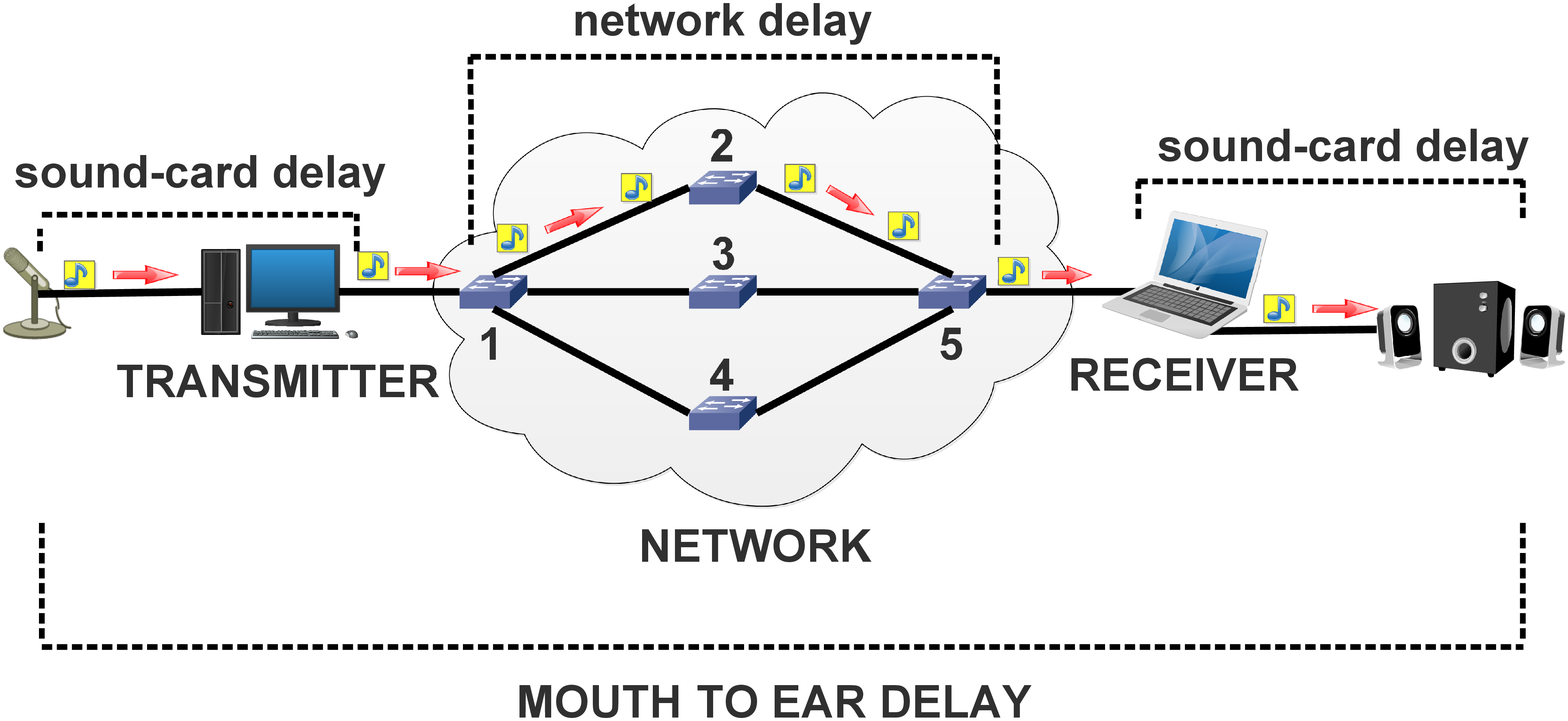}}
\caption{End-to-end delay.\label{end_to_end_delay}}
\end{figure}

From the above description, the total $mouth-to-ear$ (see Figure \ref{end_to_end_delay}) delay in an NMP can be expressed via the following Equation:
\begin{equation} \label{eq:1}
d_{mouth-to-ear}=d_{sound-trans} +d_{proc-trans} + d_{network} + d_{sound-rec} +d_{proc-rec}
\end{equation}
where $d_{mouth-to-ear}$ denotes the $mouth-to-ear$ latency (or Over-all One-way Source-to-Ear (OOSE) delay \cite{rottondi2016overview}), $d_{sound-trans}$ is the delay inserted by the transmitter's sound-card, $d_{network}$ is the delay added due to the streaming process through the network and $d_{sound-rec}$ is the delay inserted by the receiver's sound-card. $d_{proc-trans}$ and $d_{proc-rec}$ describe the delay inserted due to the audio processing and the encoding/decoding in the transmitter/receiver side.
In more detail, the end-to-end delay in NMP systems involves many factors that affect it \cite{rottondi2016overview}. These factors are:
\begin{itemize}\label{full_expression}
\item the propagation delay for the sound in the air between the audio source and the microphone,
\item the transformation process from sonic wave to electric signal via the microphone,
\item the audio signal transmission via the microphone to the transmitter computer,
\item the analog to digital conversion, combined with additional encoding process, in the transmitter's sound-card using drivers,
\item the preparation stage for the audio packetization before the transmission via the Internet,
\item the propagation delay using the Internet (network delay),
\item the depacketization and decoding process in the receiver side,
\item the application buffering process in the receiver side,
\item the driver buffering using the receiver's sound-card and the digital to analog conversion process,
\item the signal transmission using the audio output connector,
\item the transformation process from electrical signal to sonic wave using the audio output devices, such as headphones or loudspeakers,
\item the propagation delay via the loudspeakers in the air to the receiver's ear,
\end{itemize}
In our approach we focus on the audio propagation delay via the Internet (network delay) and the delay created due to audio processing and packetization/depacketization process (blocking delay). The other types of delay described in List \ref{full_expression} are negligible. Due to the high delay inserted by the encoding/decoding process, uncompressed audio is transmitted. For this reason, Equation (\ref{eq:1}) becomes:
\begin{equation} \label{eq:2}
d_{mouth-to-ear}=d_{sound-trans} + d_{network} + d_{sound-rec}
\end{equation}
In cases where the transmitter-receiver pair uses sound-cards with similar specifications regarding to reading/recording processes (assume that $d_{sound-trans}$=$d_{sound-rec}$=$d_{sound}$), Equation (\ref{eq:2}) is transformed to:
\begin{equation}
d_{mouth-to-ear}=2 \times d_{sound} + d_{network}
\label{eq:3}
\end{equation}
Equation (\ref{eq:3}) expresses the $mouth-to-ear$ delay evaluation in NMP systems between a pair of transmitter and receiver that communicate using the network. In each side (transmitter or receiver), the delay created by the audio hardware during the capturing process is called blocking delay. It is a key feature for each sound-card performance \cite{carot2007network}. Equation (\ref{eq:4}) describes the blocking delay evaluation process.
\begin{equation}
d_{blocking-delay}= \frac{frame\ size}{sampling\ rate} + d_{0}
\label{eq:4}
\end{equation}
Considering Equation (\ref{eq:4}), three factors affect the blocking delay: sampling rate, frame size and $d_{0}$. Frame size expresses the packet size that sound-cards produce as output per hardware clock tick. Sampling rate denotes the number of samples that the sound-card acquires per second. The last term, $d_{0}$ regards the constant delay inserted by the sound-card and is related to the hardware quality. In Equation \ref{eq:4}, $d_{blocking-delay}$ equals to $d_{sound}$ denoted in Equation \ref{eq:3}.

Table 1 shows that in order to achieve a decrease in blocking delay, the fraction between the frame size and the sampling rate should decrease, which is feasible via frame size decrease or sampling rate increase. In general, decreased blocking delay can benefit the NMP performance in cases that the network delay increases resulting in over-EPT end-to-end delay.
\begin{center}
\begin{table}[!ht]%
\centering
\caption{Example sound-card latency.\label{blocking_delay_matrix}}%
\begin{tabular*}{300pt}{@{\extracolsep\fill}lcccc@{\extracolsep\fill}}
\toprule
\textbf{Sampling rate (Hz)} & \textbf{Frame size (samples)}  & \textbf{Blocking delay (ms)}\\
\midrule
2048 & 48000 & 42.6\tnote\\
1024 & 48000 & 21.3   \\
512 & 48000 & 10.6   \\
256 & 48000 & 5.3  \\
128 & 48000 & 2.6  \\
64 & 48000 & 1.3  \\
\bottomrule
\end{tabular*}
\end{table}
\end{center}

Apart from the audio perspective, the network is a critical factor for the end-to-end delay in NMPs. Network performance affects the playing strategies among the musicians that participate in an NMP event. In the research community, there are 3 approaches for compensating for the end-to-end delay \cite{carot2007network}. The first approach introduces realistic interaction between the musicians, without considering the delay constraints. This approach is the only way to provide high quality in NMPs used by professional musicians. The second approach is based on the master-slave model: in each pair of players, one player has the leading role and decides the music timing, which the other player has to follow and adjust his playing. In this case, the network delays are affordable up to 100-200 milliseconds self-delay tolerance threshold \cite{barbosa2011influence}. The third technique relies on adding artificial delay in each audio stream. This delay is equal to the audio Round-Trip Time in the NMP system. This approach requires a metronome for perfect synchronicity at both the transmitter and the receiver. However, this technique is vulnerable to clock timing issues between local clocks.

About the network behavior, in delay-sensitive applications such as NMPs, we need to embed logic into the network for instant detecting and solving problems during the NMP event. These problems could be traffic congestion, link failure, natural disaster etc. For this reason, we look for an approach that can cope with such problems. In our proposed approach introduced in section \ref{sec:system_design}, we introduce a new technology in computer networks area called Software Defined Networking (SDN). It minimizes the rigidity and static profile in conventional networks. It's main target is increasing the degree of network flexibility and adaptability to each user demands. The importance of SDN is strengthened by the comparison with traditional network architectures.

In conventional networks, all forwarding devices participating in transmission process (e.g switches and routers) have similar features refering to design and functionality \cite{valdivieso2014sdn}. The major concept is that there is hardware specialized in packet processing that composes the control plane and the hardware is ruled by an operating system that collects information from the hardware and runs software application, which is called data plane. The software application is a program with thousands lines of code that define the network behavior and follow the rules for each protocol, defined by the appropriate RFCs or vendor specification manuals. The major disadvantage of this process is that the code is not accessible by the network administrator. The administrator is limited to modify the network behavior using low level commands via the command line interfaces (CLI). This fact requires high programming skills from the network administrator, as well as deep knowledge of the source code for the control plane.

Additionally, in various protocols each node exchanges information with the neighboring nodes in order to decide the next hop. These protocols lack a centralized entity, which is aware of the whole network condition and can make routing decisions. Also, when a rerouting decision should be taken, the administrator should modify the software in each device, which makes the rerouting process complex and time wasting. In recent decades,  new technological trends have arisen. In computer networks field, terms such as VLAN, IPV6 and QoS have appeared in human life. This fact indicates that computer network area should be able to adopt the recent changes. The update process in every operating system is feasible when the software is separated from the hardware part. Especially, update process or experiments is a complex task since there are compatibility problems among devices from different vendors.

The limitation for the researchers to experiment in real networking hardware devices and the required time, effort and cost leads to creating new re-programmable network devices in order to reduce the total cost for the research community. Additionally, some of the primary objectives in the network research community, such as path management, traffic engineering and recovery methods perform great computational complexity due to the growth of the Internet traffic in recent years. An approach towards coping with these difficulties is given via the Network Function Virtualization (NFV) \cite{koumaras2016service}. NFV describes the technology virtualization methods that are used to provide fundamental network functionalities and by connecting them, communication services are feasible. The NFV approach can be applied for experimental purposes instead of using the real infrastructure, which is expensive. The interconnection among functions that use the NFV provides to users end-to-end connectivity, creating a Virtual Network Function as-a-Service (VNFaaS) \cite{bueno2013opennaas}. Since the NFV functionalities are programmable, they can be used for providing certain Quality of Service to the users.

The demand for increased flexibility in network management that is offered by NFV is satisfied by the Software Defined Networking (SDN), that has been embraced at large in industrial and research domain, promising agility and optimization in network management. SDN is a powerful concept to increase flexibility and adaptability in recent communication networks \cite{durner2015performance,lin_enabling_2014}.

Software Defined Networking inserts major changes in current network architectures. First, it proposes the decoupling between the control plane and the data plane \cite{valdivieso_caraguay_sdn:_2014}. This allows each of the two separated fields to be developed independently, enabling evolution and innovation in both planes. The control plane logic is removed from the forwarding devices (switches and routers) that participate in the network and is incorporated into an external entity that instructs them and is denoted as the SDN controller. The communication between the SDN Controller and the switches is feasible via the OpenFlow protocol \cite{limoncelli_openflow:_2012}. OpenFlow allows the communication between the data plane that the switches contain and the control plane that is located in the controller side. This communication takes place via a secure channel that the controller uses to send OpenFlow messages to the switches and receive from them. The messages are translated into flow rules, that are stored into the forwarding tables at the switches, defining their behavior and the routing policies. Instead of a single controller, for large scale networks, a cluster of SDN controllers can be used for better network management \cite{sharma_demonstrating_2014}. Single controller architecture is simpler and cheaper but centralized approach raises problems regarding to scalability. As the number of switches increases, relying on a single controller is not secure for many reasons: first, instructions from the SDN controller are passed to the switches as messages. The amount of the control messages increases with the number of switches. Additionally, if the network diameter increases extremely, some distant switches will have longer setup delay compared with the nearest switches. Finally, the controller's processing power bounds the Software Defined Networking performance. Large number of switches can cause excessive setup times, treating the network's secure performance \cite{fernandez_comparing_2013}. The SDN controller can be treated as the Network operating system (NOX) that allows the management scripts to be executed over high level abstractions for defining the network behavior \cite{kwangtae_jeong_qos-aware_2012}.

\section{System design}
\label{sec:system_design}

\begin{figure}[!b]
\centerline{\includegraphics[width=0.8 \textwidth]{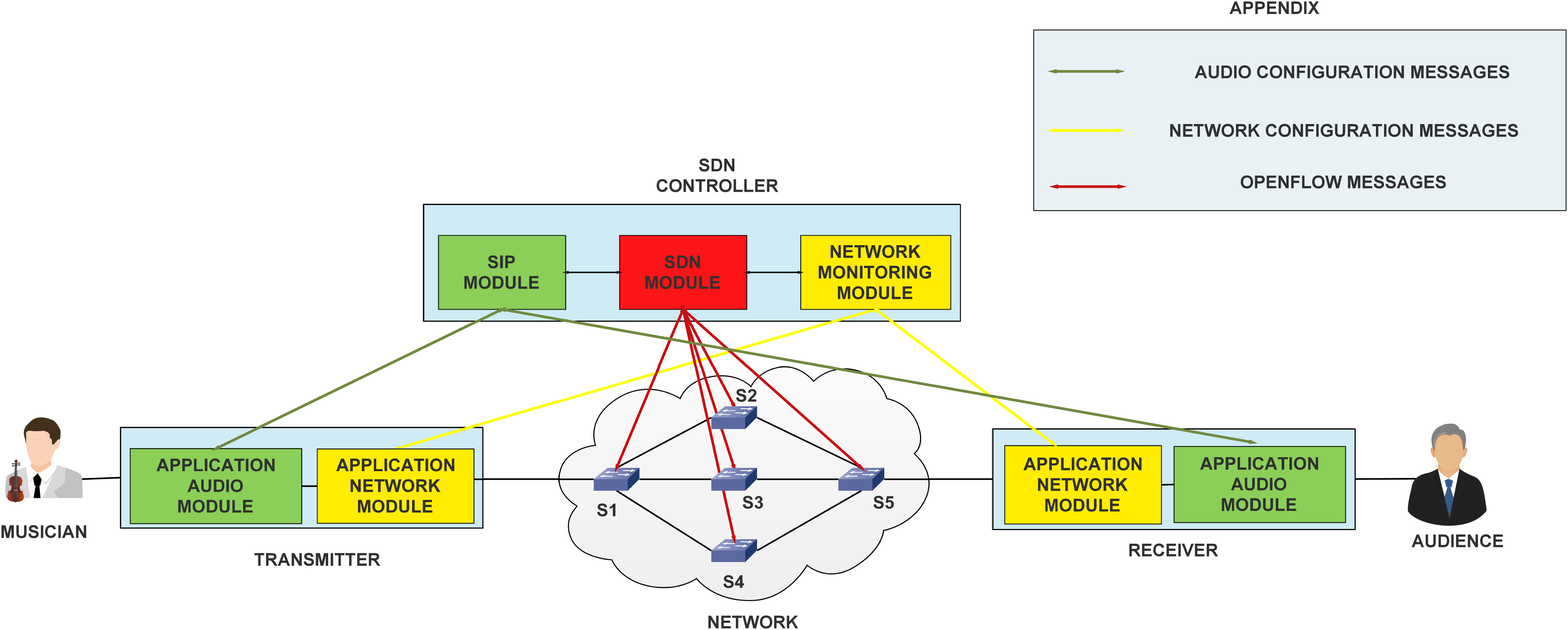}}
\caption{System architecture.\label{system_architecture}}
\end{figure}

The proposed architecture approaches the NMP systems introducing the collaboration between the application and the network. The network behavior is defined using the SDN technology  \cite{akyildiz_roadmap_2014}, which is described in section \ref{sec:background}. The major components of the proposed architecture are shown in Figure ~\ref{system_architecture}. They include:

\begin{itemize}
\label{abstract_components_list}
\item the Transmitter,
\item the Receiver,
\item the Network,
\item the SDN controller
\end{itemize}

In our approach, the controller has extended functionality compared to conventional SDN applications. In more detail, the controller can interact dynamically with the application side. This communication includes replying to application for path requests with certain delay and jitter. The controller also monitors all available paths in real-time to discover traffic congestion. In such a case, it dynamically decides to reroute the audio flows to an alternative path which performs with less delay. Also, if there are no available paths that satisfy application's requirements, the SDN controller informs the application side to modify the audio processing parameters, aiming to reduce the audio processing (blocking) delay. This can result in significant end-to-end delay decrease, which is crucial in ultra-low delay sensitive applications, like the NMP systems. The target of this process is to overcome network delay increase by audio processing modification.

The role and the functionality that each component performs will be discussed in the next subsections.
\subsection{Transmitter}

In the proposed implementation schema, the Transmitter component denotes the entities that generate and transmit audio via the network. In the NMP case, the Transmitter component represents any musician that participates in the teleorchestra live performance. The Transmitter component contains two modules: the Application Audio module and the Application Network module. The Application Audio module is responsible for providing information about the sound-card's performance to the SDN controller in order to create the audio profile for each user. The audio profile contains the sound-card performance in terms of blocking delay for various sampling rate and frame size combinations. The Application Audio module also is assigned to capture audio from the sound-card in order to transmit it via the network. In case of traffic congestion, the Application Audio module receives the instructions for audio processing pattern modification in order to tolerate with further network delay increase. In general, the signaling and control duties of the Application Audio module are similar to Session Initiation Protocol (SIP) operation, where users negotiate for the audio processing configuration set during the transmission process.

The Transmitter component contains also the Application Network module. This module is responsible for the audio transmission via the network from the transmitter to the receiver. Also, it collaborates with the SDN controller in participating in the network delay monitoring process. The Transmitter evaluates the network delay in real time and this information is stored in the SDN controller. Using this information, the SDN controller has an accurate view of the end-to-end delay for the transmitted audio signal, combining the information about the network delay and the user's audio profile.

\subsection{Receiver}
The Receiver component represents all users in the NMP system that receive the audio via the network. These users can be either the audience who receives the final audio signal transmitted via the NMP system or the musicians that participate in the live music performance and they receive particular audio plays from other musicians in order to stay synchronized. The Receiver component uses the same modules as the Transmitter component but in a different way. In more detail, the Receiver uses the Application Audio module for informing the SDN controller about the audio profile of the receiver, and also is responsible for receiving and playing the received audio signal. Similar to the Transmitter component, the Application Audio module also receives the notification by the SDN controller in case of traffic congestion in order to modify the audio processing pattern. Finally, the Receiver uses the Application Network module in the network delay monitoring process, that will be discussed later.

\subsection{Network}
The Network represent the network infrastructure that is used to provide connectivity between the users for the successful audio transmission. The Network denotes the real world Internet, that is used in daily life by billions of users for data transmission. The Network consists of forwarding devices (e.g. switches) that use OpenFlow protocol in order to be able to receive instructions by the SDN controller. The network topology is created by linking the OpenFlow switches between them and also with the end hosts.

\subsection{SDN controller}
\label{sub:sdn_controller}
The SDN controller is the main entity in our architecture. Apart from the conventional duties that it has for taking routing decisions and installing flow rules, it is also capable of interacting with the application side in order to inform it about the network condition. This communication acts as a warning in case of traffic congestion, enabling a timely audio re-configuration.

The proposed SDN controller component contains three modules. The first is called SIP module. It is assigned to collect the audio profiles from each user when he registers in the NMP application. This information is stored in a database for each user. The SIP module is triggered also in case of traffic congestion in order to instruct the application to modify its audio processing pattern. The second module is the SDN module. This module is responsible for the path setup via installing appropriate flow rules into the network switches that form the path between each pair transmitter-receiver. The communication between the SDN module and the switches is feasible via OpenFlow protocol ~\cite{akyildiz_roadmap_2014}. The last module of the SDN controller is called Network Monitoring module. The Network Monitoring module's duty is the network delay monitoring. It executes real-time measurements for the network delay over time for each path and keeps this information in a database. This information is useful for the SDN controller in order to have an accurate estimation of the real-time end-to-end delay, and reroute the audio packets to another path in case of traffic congestion.

\subsection{Real-time network delay monitoring}
The Network Monitoring module is responsible for monitoring the network delay for each path in the network. Many SDN-oriented approaches that require real time monitoring of the network evaluate the network workload by requesting statistics from the switches that form the network. OpenFlow allows for extracting data rate statistics from switches using appropriate queries. This process is called polling. The inter-arrival time between consecutive statistics requests is called polling period \cite{tomovic2014sdn}. For low polling period values, switches are overloaded with requests in a short period of time, causing problems in their performance. On the other hand, for high polling period values, the computational workload over time will decrease but the estimation for the network delay will not be accurate, since the time period between two consecutive statistics requests is large. Thus, there is a tradeoff in selecting the appropriate value for the polling period.

\begin{figure}[!ht]
\centerline{\includegraphics[width=0.5 \textwidth]{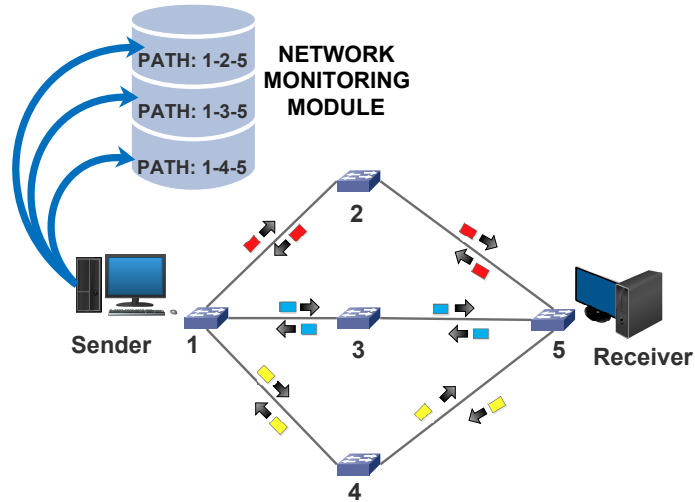}}
\caption{Network delay monitoring approach.\label{monitoring_delay}}
\end{figure}

In our proposed network monitoring approach, the switches are static in the network monitoring process, which reduces their workload. The required complexity is moved to the end hosts that have more resources than the OpenFlow switches. Each pair of the transmitter and the receiver starts exchanging UDP packets with a specific format over time. The transmitting node  sends the UDP packets over each path and the receiver has the role of an echo server that replies back to these packets, following the inverse path. The Round Trip Time of each received packet at the transmitter side represents the current network delay for each path. The polling period in our architecture is one second. Each time a packet is received by the transmitter, the evaluated RTT result is stored into the database, updating the view for the network condition that the SDN controller has for each path. Figure \ref{monitoring_delay} depicts the network delay monitoring process. The arrows show the direction of the packets, while they travel through the network.

\subsection{Audio profiling}
\begin{figure}[t]
\centerline{\includegraphics[width=0.6 \textwidth]{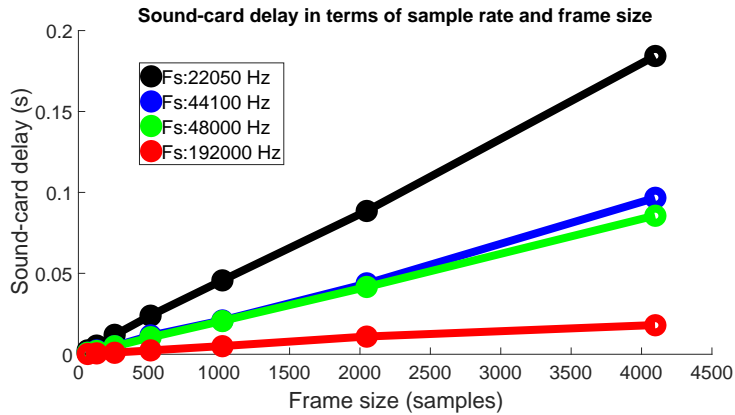}}
\caption{Sound-card blocking delay.\label{blocking_delay}}
\end{figure}

When a new user joins in the NMP application, our architecture via the SIP module should create the audio profile for this user. The term audio profile denotes the user sound-card performance in terms of blocking delay for various sampling rate and frame size combinations. This information is useful for the SDN controller in order to evaluate accurately the end-to-end delay as the summary of the network delay and the blocking delay over time. All user profiles are stored into a database in the SDN controller side. Based on the audio profile for each user, the SIP module can instruct users to change their audio processing parameters (sampling rate and frame size) in order to cope with further network delay increase. This is feasible by decreasing the frame size and the sampling rate, based on Equation (\ref{eq:4}). For neighboring sampling rate and frame size combinations, the quality of the transmitted signal has no significant changes otherwise the quality will temporarily decrease until the network congestion is handled. Figure \ref{blocking_delay} depicts a sample of audio profile for a sound-card used in our experiments. The plots confirm the blocking delay behavior in terms of sampling rate and frame size values.

\subsection{Path setup and rerouting decisions}
\label{sub:path_monitor}
After creating the audio profile for each user, our system can serve path requests for any transmitter-receiver pair. The transmitting node declares interest to transmit audio to a receiver node. The request sent by the user, apart from the destination node, contains also the delay and jitter requirements. The SDN controller examines the path request and given the audio profile of each user and the network delay evaluated by the network delay monitoring, chooses the appropriate path that satisfies the user requirements. After this stage, the audio transmission process starts and audio packets travel through the network. The SDN controller, apart from the used path, has also estimated additional paths for redundancy in case of traffic congestion or link failure.

As the transmission process continues, the SDN controller monitors all available paths assigned to each separate audio flow. In case that the current path shows higher delay than the delay in the available paths, the SDN module takes a rerouting decision, installing flow rules for the selected path using OpenFlow. This approach guarantees that the current path is always the path with the lowest delay among the available paths. In order to avoid continuous rerouting decisions between paths, we use a threshold for taking rerouting decisions. In case that the difference between delay in an available path and the current path is higher or equal to the threshold value, the SDN controller redirects audio flows to this path.

\subsection{Audio processing reconfiguration}
As we described in subsection \ref{sub:path_monitor}, the SDN controller monitors in real time the network delay over each available path and chooses the path that satisfies the requirements for the NMP event. The rerouting process takes place as long as there are available paths that result in below the EPT end-to-end delay. If all paths are congested, the SDN controller, using the SIP module, instructs both the transmitter and the receiver to modify their audio processing pattern, switching to a combination with decreased blocking delay. This decision can possibly cause a drop in audio quality. For this reason, the SIP module chooses the combination that offers below EPT end-to-end delay combined with the minimum possible quality change.
\section{Evaluation}
\label{sec:evaluation}
We implemented an emulation scenario to test the attained NMP quality of service using the proposed architecture of application-network collaboration. Our goal is to emphasize on how the interaction between the application side and the network can benefit in terms of robustness against network changes.

\subsection{Experimental Scenario}

\begin{figure}[t]
\centerline{\includegraphics[width=0.52 \textwidth]{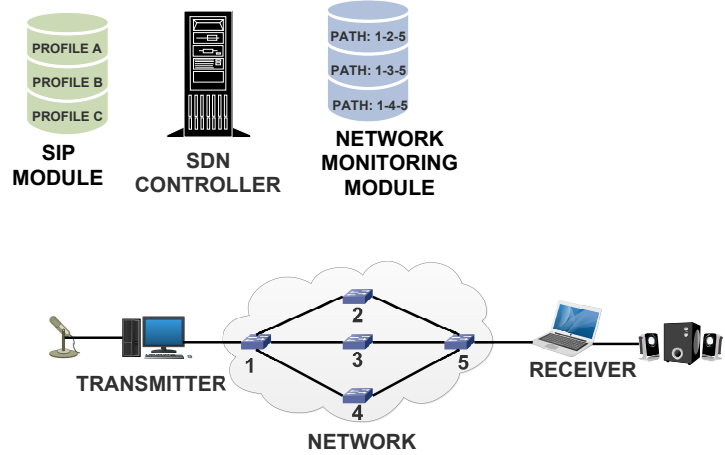}}
\caption{Experimental setup.\label{topo}}
\end{figure}

The network topology for our experiments is implemented using Mininet \cite{sharma2014mininet}, a Python-based network emulator that supports the OpenFlow protocol and allows SDN principles. We tested various topologies and the one that we present consists of five OpenFlow switches, that form the paths between the transmitter and the receiver depicted in Figure \ref{topo}. For the SDN controller, we used POX \cite{kaur2014network}, a Python-based single-threaded SDN controller, which is widely used for research experiments. Audio acquiring, processing and streaming are implemented using the Mathworks Simulink software in the transmitter and receiver side. For increasing the delay in the paths, we used the Netem traffic control tool ~\cite{hemminger2005network}. In our experiments we allow switching between two audio configuration modes. The default has 22050 Hz as sampling rate and 128 samples as frame size. The second audio mode, that is selected in case of global network congestion refers to 44100 Hz for sampling rate and 64 samples frame size. The blocking delay for the two audio modes is described in Table 2.
\subsection{Creating audio profiles}
The first task for our system is creating an audio profile for each user. The audio profile depends on the sound-cards used. In our case, we used similar sound-cards in the experimental setup. We evaluated the blocking delay for various sampling rate and frame size combinations. The blocking delay denotes the delay between the initial audio signal that inserts into the sound-card and the signal that is played. For this reason, we used an audio source that sends audio signal to the sound-card input and we evaluated the cross-correlation between the signal that enters the sound-card and the signal that exits from the sound-card. This result equals to the blocking delay for the sound-card. Table 2 shows the results of this process. Figure \ref{blocking_delay} is a graphical representation of these results.

\begin{center} \label{evaluated_blocking_delay}
\begin{table}[!ht]%
\centering
\caption{Evaluated sound-card delay.\label{tab2}}%
\begin{tabular*}{300pt}{@{\extracolsep\fill}lcccc@{\extracolsep\fill}}
\toprule
\textbf{Frame size (samples)} & \textbf{Sampling rate (Hz)}  & \textbf{Blocking delay (ms)}\\
\midrule
64 & 22050 & 2.29 \\ \hline
128 & 22050 & 5.41 \\ \hline
256 & 22050 & 11.84 \\ \hline
512 & 22050 & 23.95 \\ \hline
64 & 44100 & 1.05   \\ \hline
128 & 44100 & 2.52  \\ \hline
256 & 44100 & 5.07  \\ \hline
512 & 44100 & 11.53 \\ \hline
\bottomrule
\end{tabular*}
\end{table}
\end{center}

\subsection{Emulation results}
\label{emulation_scenario}
In order to check the performance of our approach, we start to increase the network delay in each path and examine the system behavior. We present the results from three experiments. In each experiment, we increase the number of the supported audio configuration modes so in case of traffic congestion, the application side has more options to choose for the re-configuration.  For experiment 1, we use two audio modes, for experiment 2 we use three audio modes and finally for the last experiment we use eight audio processing modes. In all experiments the transmitter and the receiver participate in a typical NMP system, so the only constraint is that the end-to-end delay should be up to 25 ms (EPT). In all experiments the Transmitter initially requests a path from the SDN controller in order to transmit audio to the Receiver. The SDN controller chooses the path with the lowest network delay and the NMP event starts. During the transmission, we add delay to the currently used path. The SDN controller monitors all available paths and in case that another path has lower delay than the current path, it reroutes audio to this path. This process continues as long as not all paths are congested. Otherwise, the SDN controller interacts with the application side in order to modify audio processing, decreasing blocking delay. In the first experiment we allow only two audio processing modes so the SDN controller  has only one option in case of network congestion.

\begin{figure}[!b]
\centerline{\includegraphics[width=1 \textwidth]{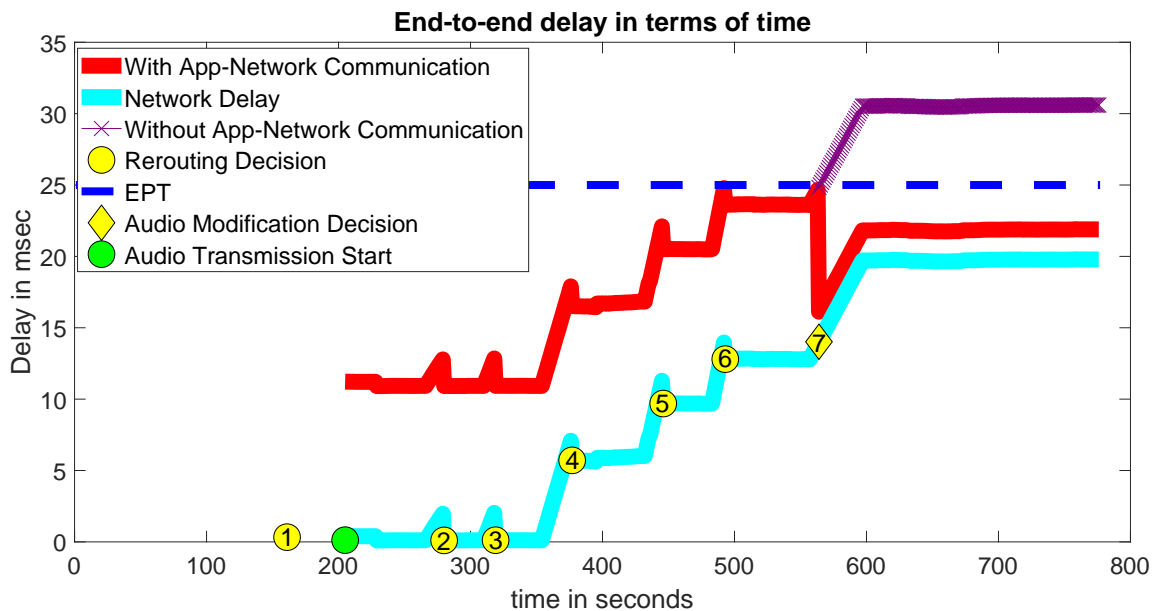}}
\caption{Blocking and end-to-end delay vs time for experiment 1.\label{evaluated_end_to_end_delay}}
\end{figure}

Table 3 describes the time that rerouting events and audio processing modifications happen during the first experiment. Each event is assigned an Event ID in order to be represented in Figure \ref{evaluated_end_to_end_delay}.

\begin{center} \label{transitions}
\begin{table}[!ht]%
\centering
\caption{Transition table.\label{tab2}}%
\begin{tabular*}{300pt}{@{\extracolsep\fill}lcccc@{\extracolsep\fill}}
\toprule
\textbf{Time (s)} &\textbf{Event ID}  & \textbf{Current Path}  & \textbf{Next Path} & \textbf{Action}\\
\midrule
161 & 1 & - & 1-3-5 & Path assignment\\
280 & 2 & 1-3-5 & 1-4-5 & Rerouting\\
319 & 3 & 1-4-5 & 1-2-5 & Rerouting\\
377 & 4 & 1-2-5 & 1-3-5 & Rerouting\\
446 & 5 & 1-3-5 & 1-4-5 & Rerouting\\
493 & 6 & 1-4-5 & 1-2-5 & Rerouting\\
564 & 7 & 1-2-5 & 1-2-5 & Audio modification\\
\bottomrule
\end{tabular*}
\end{table}
\end{center}

As we mentioned, the major performance metric for an NMP system is the end-to-end delay. For this reason, Figure \ref{evaluated_end_to_end_delay} depicts on aggregate all useful information for this experiment.

In Figure \ref{evaluated_end_to_end_delay}, the network delay (dark blue solid line), the end-to-end delay (red line) and the overall blocking delay created by the two sound-cards (in the transmitter and the receiver side) for both audio processing modes are shown. At t=161 s, the transmitter sends an path request to the SDN controller. The SDN controller chooses $1-3-5$ among the available paths and informs both the transmitter and the receiver about the recommended audio configuration set. At t=200s, the users start using the NMP application so the audio transmission process starts using audio configuration set 1 with overall blocking delay (for the two sound-cards) equal to 10.82 ms (5.41 ms for each sound-card). The delay values are described in Table 2 in detail. The audio transmission start is represented using the triangle marker. During the live music performance, we add delay in each path using the NETEM traffic control tool and we check our system's behavior. As explained in subsection \ref{sub:path_monitor}, the Network Monitoring module in the SDN controller side monitors all available paths and in case that the difference between an available path and the current path is greater or equal to the threshold value (in our experiments the threshold value equals to 2 ms), the SDN module redirects the audio flows to this path by installing the appropriate flow rules in the switches that participate in the new path. For this reason, at t=280 s, the SDN module decides to reroute the audio packets to path $1-4-5$. This process continues until there are paths whose delay results in below-EPT end-to-end delay values. The cycle markers in the figure denote all the rerouting decisions that are taken during the NMP event containing the corresponding Event IDs denoted in Table 3.

In order to demonstrate the improvement in the NMP quality of service when collaboration between the application and the network is allowed, we congest all available paths using NETEM. As a consequence, at t=564 s, the SDN controller recognizes that the end-to-end delay tends to exceed the EPT constraint so it instructs the audio processing pattern modification for the transmitter and the receiver. The SIP module recommends an audio processing mode with 44100 Hz sampling rate and 64 samples frame size. The new audio processing pattern shows overall blocking delay equal to 2.1063 ms (1.0532 ms for each sound-card), which is lower that the initial blocking delay. Reducing the blocking delay can deal with further network delay increase, which increases the NMP resistance against the traffic congestion problem. In the same figure, we show that the end-to-end delay without collaboration between  the application and the network would reach above 30 ms (at t=780 s). However, after the interaction between the application and the network, the end-to-end delay reaches 21.29 ms, so the gain from this operation equals to 8.71 ms, without violating the EPT constraint.

We quantify the improvement that our approach provides to the NMP application compared with the scenario that the application side does not interact with the network. For this reason, we used two indicators. The gain in end-to-end delay terms that interaction between the transmitter and the receiver offered in percentage scale is evaluated as:
\vspace{5mm}
\begin{equation} \label{gain eq1}
gain = \frac{end-to-end\ delay_{without\ interaction}-end-to-end\ delay _{with\ interaction}}{end-to-end\ delay\ _{without\ interaction}} *100 \%
\end{equation}
\vspace{5mm}

Also we can define gain metric referring exclusively to blocking delay which is evaluated as
\vspace{5mm}
\begin{equation} \label{gain audio}
gain_{audio} = \frac{total\ blocking\ delay _{initial\ mode} - total\ blocking\ delay _{new\ mode}}{total\ blocking\ delay _{initial\ mode}} *100 \%
\end{equation}

\begin{figure}[t]
\centerline{\includegraphics[width=0.8 \textwidth]{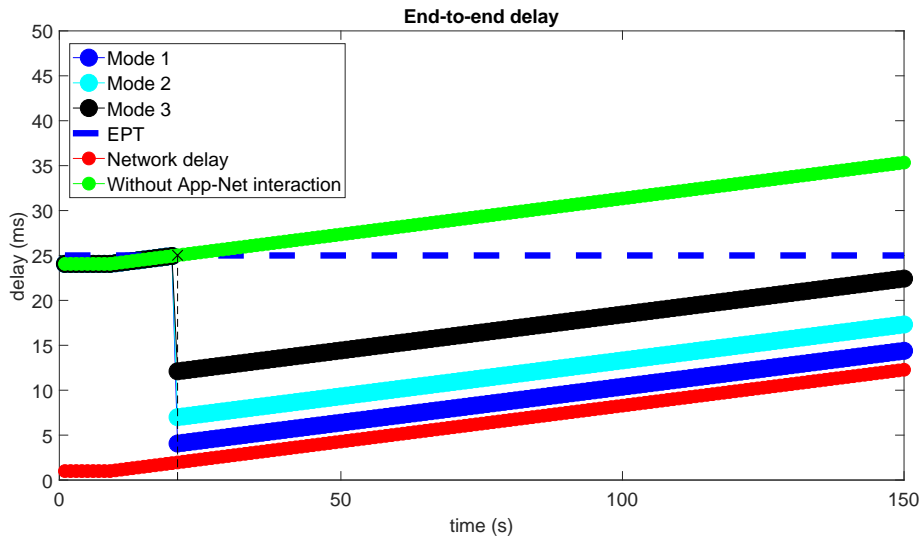}}
\caption{End-to-end delay vs time for experiment 2.\label{simulations comparison}}
\end{figure}

The gain value in Equation (\ref{gain eq1}) shows the degree that the end-to-end delay is reduced compared to the case that no audio modification is decided. On the other hand, the gain value from Equation (\ref{gain audio}) shows the degree that exclusively the blocking delay is reduced due to audio modification in a mode with reduced audio latency. In the second experiment, we tried more combinations for audio configuration parameters than in the previous experiment. The initial audio mode  refers to sampling rate equal to 44100 Hz and frame size equal to 512 samples and the candidate audio modes for re-configuration can be either with sampling rate equal to 44100 Hz and frame size equal to 64 samples (Mode 1), 128 samples (Mode 2) or 256 samples (Mode 3), the gain referring to the end-to-end delay and the blocking delay in each case is shown in Table 4. In this experiment, we increased network delay 0.08 ms per second and we consider that the network delay is equal to 1 ms second for time period until t=10 seconds. Figure \ref{simulations comparison} shows the delay behavior for each audio configuration mode. The audio configuration is decided at t=21 seconds. The values for the blocking delay in each audio mode are described in Table \ref{blocking_delay_matrix}.
\begin{center} \label{gain_matrix}
\begin{table}[!ht]%
\centering
\caption{Gain results.\label{tab2}}%
\begin{tabular*}{300pt}{@{\extracolsep\fill}lcccc@{\extracolsep\fill}}
\toprule
\textbf{Sampling rate (Hz)}&\textbf{Frame size (samples)} &\textbf{Gain(\%)} &\textbf{Gain audio(\%)}\\
\midrule
44100 & 64 & 59.29 & 90.8\\
44100 & 128 & 51 & 78.16\\
44100 & 256 & 36.55 & 56\\
\bottomrule
\end{tabular*}
\end{table}
\end{center}

\begin{figure}[!ht]
\centerline{\includegraphics[width=0.8 \textwidth]{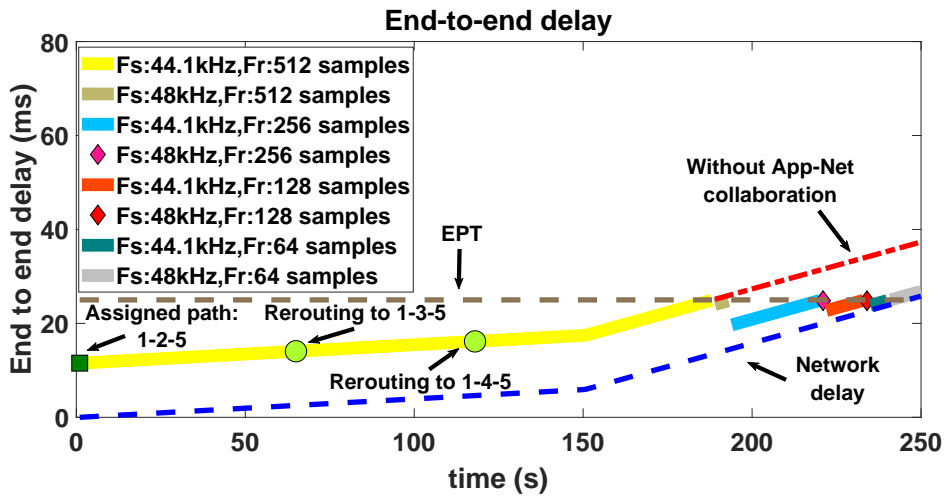}}
\caption{End-to-end delay vs time for experiment 3.\label{sim2}}
\end{figure}

In experiment 3 we use the same topology but contains more audio processing combinations than the previous experiments but we have eight possible audio configuration modes in case of network congestion. The experiment results are depicted in Figure \ref{sim2}. Assume that the Receiver node declares interest for receiving audio from the Transmitter node. There are three available paths connecting the two nodes in the network. Initially, the SDN module assigns path $1-2-5$ for the audio transmission with $F_s$ equal to $44.1$ kHz and $F_r$ set to $512$ samples. As we increase the network delay in the path by 0.04 ms per second until t=150 s and then by 0.2 ms per second, the SDN module reroutes audio flows to the remaining paths $1-3-5$ (at $t=65$ s) and $1-4-5$ (at $t=118$ s).  At $t=189$ s, the SIP module informs the NMP application to switch to an audio mode with lower blocking delay ($F_s$ equal to $48$ KHz and unaltered $F_r$). Assuming that the network delay increases further, and in order to keep the end-to-end delay below EPT constraint, at $t=194$ s, the NMP modifies $F_r$ to $256$ samples after interacting with the SIP module. This process takes place during the experimental setup for various audio modes. When the network delay reaches high levels, our system results in best effort delay (after $t=241$ s), given the available audio modes and the network condition. Comparing our method with the case that application and network could not interact, we achieve delay improvement by $29.6\%$ in the end-to-end delay.

\subsection{Discussion}
In subsection \ref{emulation_scenario}, we applied our approach for communication between the application and the network using software for emulating network topologies called Mininet. This software allows to add certain amount of delay in each interface in order to check our system's performance for various delay conditions. On the Internet, the delay in each link between network devices (switches and routers) is a function of various nodes. The first and most important factor is the geographical distance between two nodes. In \cite{singla2014internet}, researchers mention that the propagation delay between two nodes $n_1$ and $n_2$ is equal to $\frac{d(n_1,n_2)}{(c * \frac{2}{3})}$ in an optical fiber, where $c$ denotes the speed of light and $d(n_1,n_2)$ denotes the geographical distance between the two nodes. The propagation delay is the minimum delay between two nodes, since we do not consider traffic congestion in the network.

In Networked Music Performance systems, as described in section \ref{sec:introduction}, the maximum affordable end-to-end delay for audio packets traveling between two nodes should be less that 25 milliseconds. For this reason, the maximum distance that audio packets should travel in order to satisfy the EPT constraint is about 5000 kilometers on aggregate, based exclusively on the geographical distance and ignoring the network condition. Apart from the distance, another crucial factor for the network performance is traffic congestion. Due to the traffic congestion, the network forwarding devices such as switches and routers cannot cope with heavy traffic via the Internet, increasing the overall network delay \cite{bozkurt2017internet}. Additionally, link failures and re-transmission process also result in higher network delay communication.

\begin{figure}[!h]
\centerline{\includegraphics[width= \textwidth]{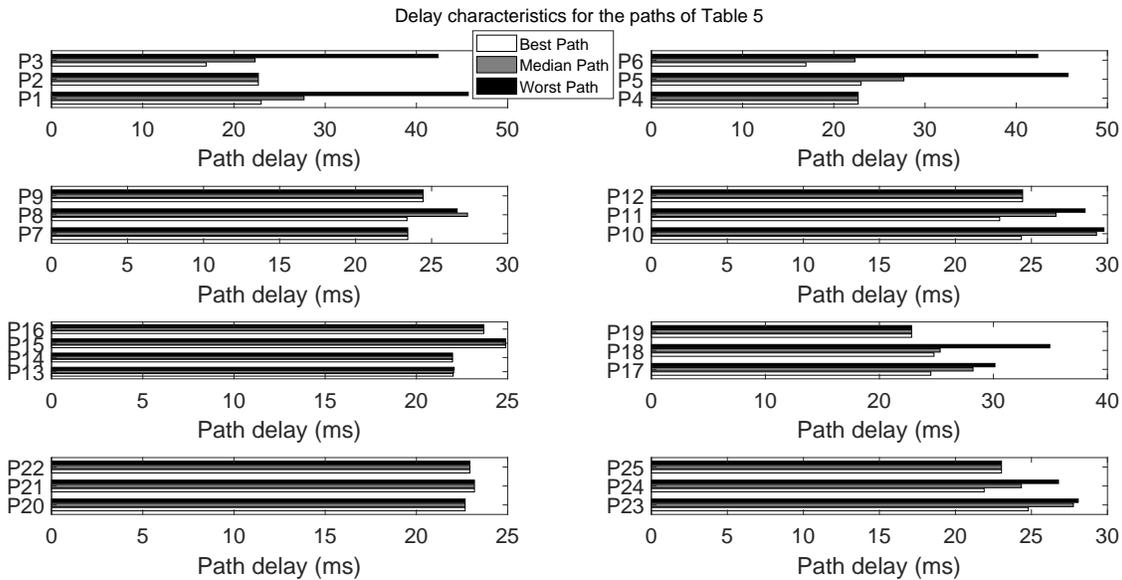}}
\caption{Minimum (white), median (grey) and maximum (black) end-to-end delay for each path denoted in Table 5.}
\label{manos}
\end{figure}

Using emulation software such as Mininet allows to add delay in the network topology but it does not take into account the factors described above that affect network performance in the real world. For this reason, we used real measurements using RIPE Atlas (RA) \cite{ripe} nodes (i.e., probes and anchors), a globally distributed measurement infrastructure located in many places around the world, consisting of end-host devices, capable of executing different types of data-plane measurements, such as latency. The data set we used for our measurements is described in \cite{kotronis2017shortcuts}. We processed the measurements and we focused on triplets of cities that could form paths that satisfy the EPT constraint and could be used for NMP purposes. In fact, we checked for pairs of cities in different countries that could participate in a Networked Music Performance using a relay node located in another city. Some of the results we found are described in Table 5.
\begin{center} \label{real_results_table}
\begin{table}[!h]%
\centering
\caption{Available cities for NMP.\label{tab2}}%
\begin{tabular*}{450pt}{@{\extracolsep\fill}lcccc@{\extracolsep\fill}}
\toprule
\textbf{Path ID} &\textbf{Source}  & \textbf{Relay}  & \textbf{Destination} & \textbf{Overall distance (km)}\\
\midrule
$P_1$ & Vienna (Austria) & Kosice (Slovakia) & Issy-les-Moulineaux (France) & 1756.34\\
$P_2$ & Vienna (Austria) & Poplar (UK) & Issy-les-Moulineaux (France) & 1571.69\\
$P_3$ & Vienna (Austria) & Ljubljana (Slovenia) & Issy-les-Moulineaux (France) & 1250.31\\
$P_4$ & Issy-les-Moulineaux (France) & Poplar (UK) & Vienna (Austria) & 1571.69\\
$P_5$ & Issy-les-Moulineaux (France) & Kosice (Slovakia) & Vienna (Austria) & 1756.34\\
$P_6$ & Issy-les-Moulineaux (France) & Ljubljana (Slovenia) & Vienna (Austria) & 1250.31\\
$P_7$ & Nitra (Slovakia) & Leudelange (Luxembourg) & Rotterdam (Netherlands) & 1176.07\\
$P_8$ & Nitra (Slovakia) & Ljubljana (Slovenia) & Rotterdam (Netherlands) & 1350.82\\
$P_9$ & Nitra (Slovakia) & Blackheath (UK) & Rotterdam (Netherlands) & 1655.84\\
$P_{10}$ & Riga (Latvia) & Tallinn (Estonia) & Blackheath (UK) & 2060.68\\
$P_{11}$ & Riga (Latvia) & Tampere (Finland) & Blackheath (UK) & 2328.29\\
$P_{12}$ & Riga (Latvia) & Rotterdam (Netherlands) & Blackheath (UK) & 1701.99\\
$P_{13}$ & Sandyford (Ireland) & Leudelange (Luxembourg) & Gamlingay (UK) & 1473.78\\
$P_{14}$ & Sandyford (Ireland) & Nijmegen (Netherlands) & Gamlingay (UK) & 1247.86\\
$P_{15}$ & Sandyford (Ireland) & Middelkerke (Belgium) & Gamlingay (UK) & 892.7\\
$P_{16}$ & Sandyford (Ireland) & Asnieres-sur-Seine (France) & Gamlingay (UK) & 1169.45\\
$P_{17}$ & Tampere (Finland) & Riga (Latvia) & Rotterdam (Netherlands) & 1892.82\\
$P_{18}$ & Tampere (Finland) & Tallinn (Estonia) & Rotterdam (Netherlands) & 1750.46\\
$P_{19}$ & Tampere (Finland) & Blackheath (UK) & Rotterdam (Netherlands) & 2137.46\\
$P_{20}$ & Arhus (Denmark) & Rotterdam (Netherlands) & Blackheath (UK) & 914.75\\
$P_{21}$ & Arhus (Denmark) & Alkmaar (Netherlands) & Blackheath (UK) & 877.96\\
$P_{22}$ & Arhus (Denmark) & Leudelange (Luxembourg) & Blackheath (UK) & 1262.39\\
$P_{23}$ & Blackheath (UK)) & Nitra (Slovakia) & Ljubljana (Slovenia) & 1712.03\\
$P_{24}$ & Blackheath (UK)) & Alkmaar (Netherlands) & Ljubljana (Slovenia) & 1367.57\\
$P_{25}$ & Blackheath (UK)) & Rotterdam (Netherlands) & Ljubljana (Slovenia) & 1294.64\\
\bottomrule
\end{tabular*}
\end{table}
\end{center}
For the scenarios described in Table 5, we evaluated the network delay between each source and destination city pairs using one city as relay node. Also, we evaluated the total geographical distance that packets should travel in each scenario, starting from source city towards the destination city and passing through the relay city. Additionally, in every path denoted as $P_i\ \forall i=1, \cdots , 25$ we used measurements extracted from the RIPE API \cite{ripe}. For every path we evaluated the minimum, median and maximum delay in order to build an abstract view for the network condition. These measurements are depicted in Figure \ref{manos}.
\section{Related Works}
\label{sec:related_work}
In the current section, we provide the research progress about NMPs. Due to exponential growth of Internet connectivity in recent decades, a large number of studies focus on researching patterns for network traffic shaping. This is crucial in order to cope with the huge amount of traffic that travels via the Internet and simultaneously satisfy the great demand for high quality of services (QoS) by the users. Service providers are interested in providing high QoS as well as they take into account the users' feedback in order to improve the provided QoS. Major indicators for QoS are Quality of Experience (QoE) and Mean Opinion Score (MOS).

In addition to the Internet growth, new technologies have appeared that require Internet connectivity. They exploit computer networking for multimedia delivery or communication purposes. NMPs belong to the first category. Research community has approached NMPs from two different perspectives based on the two structural components evolved in NMPs: the application side and the network. Both approaches aim to cope with end-to-end delay increase from a single perspective (audio or network oriented) but the improvement is limited since the end-to-end delay in NMPs is the summary of both perspectives. In the current section, we analyze both perspectives met in NMP related work.

\textbf{The audio-centric perspective}: From the audio perspective, many research attempts focus on the way that the audio signal is transmitted towards the receiver. The common solution followed in research studies is that each transmitter sends one or more audio streams \cite{baltas2014ultra}. Receivers apply appropriate filters in order to ignore the audio flows that they are not interested. Each receiver creates an audio profile indicating the related audio streams and based on these preferences, the receiver acquires the corresponding audio flows. This filtering process for matching the audio streams to the audio profiles increases the end-to-end delay, preventing from using this approach in the NMP systems. Based on the amount of transmitted traffic, transmission type can be either multicast or selective forwarding. Using multicast, each audio flow is transmitted to all participants. Multicast does not require a filtering stage so it does not cause a delay increase but it results in network under-utilization since all participants receive the same amount of audio traffic, without taking into account the user preferences. On the other hand, selective forwarding is based on user preferences, reducing the amount of audio traffic in the network, enabling better bandwidth management but it increases overall delay due to the filtering stage.

In the selective audio transmission scenario, the operation is assigned to a central entity called Selecting Forwarding Unit (SFU). This component collects all transmitted audio flows and based on the user profiles, forwards the appropriate audio flows to the  receivers \cite{carot2007network,carot2006network,carot2007networked,xiaoyuan_gu_network_centric_2005}. Users create their audio profiles by sending requests to the SFU. Additionally, many researchers evolve the role of SFU. In more detail, SFU is replaced by a component called Multipoint Conferencing Unit (MCU) \cite{akoumianakis_musinet_2014,xylomenos2013reduced}. The role of MCU is that it receives all audio streams from the transmitting participants, it mixes them and transmits the audio result to the receiving participants.

Another common trend in NMP system is that the Session Initiation Protocol (SIP) is used for control messages among the transmitter and the receiver side \cite{sinnreich_internet_2006}. SIP is a protocol widely used in parallel with Real Time Protocol (RTP) for initial handshaking and dynamic transmission modifications during runtime \cite{nam2014towards, nurmela2007session, ali2013session, sinnreich_internet_2006, camarillo_evaluation_2003}. Using SIP, each transmitter-receiver pair can agree to required encoding/decoding pattern before and during the transmission process.

In the specific case of NMPs, there are many projects for real time audio transmission. They can be grouped in the following categories:

\begin{itemize}
\label{nmp_classes}
\item Realistic Jam Approach (RJA)
\item Latency Accepting Approach (LAA)
\item Remote Recording Approach (RRA)
\end{itemize}

RJAs describe applications where real time live music interactions are critical so the final music result should be similar with the case musicians would be performing in the same place. This means that the delay due to the distance between them should be minimized. RJA applications target to send and receive data as quickly as possible, given a remarkable network connection. In the RJA approach belong projects such as Soundjack and eJamming. Soundjack \cite{carot2006network} is an application developed by Alexander Carot and inspired by the SoundWIRE project. Soundjack allows direct access to the sound-card and uses UDP for the audio transmission. It relies on peer-to-peer connectivity without requiring any relay node. The audio quality is a function of sound-card's quality and network quality. In order to avoid jitter cases, Soundjack allows buffer size adjustments for absorbing delay variance. Ejamming \cite{ejamming} is similar to Soundjack but it uses MIDI encoding, reducing the required bandwidth. Users can define the maximum affordable jitter above which signal is not played out. This bottleneck is called Late-Note Tolerance. Also, eJamming inserts the approach of delayed feedback \cite{barbosa2006displaced}, which adds delay to same instrument sound in order to be synchronized with the incoming sound from external participants.

The second category is called Latency Accepting Approach (LAA). This class of applications accepts network delays at level of 200ms. These applications are regarded as ultra-low delay sensitive. Well-known LAA projects are Ninjam and Quintet.net. Ninjam \cite{ninjam} is based on the assumption that due to the network delay, musicians cannot perform synchronously. In order to synchronize the transmitted audio flows, delay cancellations mechanisms are used. Quintet.net is a network performance application developed by the composer and computer musician Georg Hajdu \cite{hajdu2003quintet}. It requires centralized control for collecting and re-transmitting audio to users. This control is assigned to a component called conductor, which can change dynamically the transmission audio parameters and informs the musicians for possible problem encountered. It also supports video context. The assumption of certain network delays requires adaptation during mixing in server sides for better acoustic result.

The last category is called Remote Recording Approach (RRA). This approach regards Internet as the medium for remote recording sessions. The transmitted audio signal is "time stamped" and this allows delay cancellation or absorption when receiving. RRA applications do not support real time human to human interaction. In this category belong Digital Musician Link (DML) and VSTunnel. Digital Musician Link (DML) \cite{dml} allows dynamic negotiation between users about the audio processing parameters. Before DML operation, each user should pass the authentication process. DML provides several levels of provided service based on the fee paid during registration. The VSTunnel Plug-In \cite{vst} allows users to create or join existing sessions. These sessions can be either public or private. Users can view the ongoing sessions and join when interested. In case of audio parameter modifications, the application informs all participants in order to adjust to the new audio processing pattern.

Apart from the applications mentioned above, there are recent applications like Jacktrip \cite{caceres2010jacktrip}, Distributed Imersive Performance \cite{sawchuk2003remote}, Diamouses~\cite{alexandraki2008towards} etc. All these applications transmit uncompressed audio streams which require an excessive amount of bandwidth. For this reason, in order to support these technologies,  leased lines or high quality Internet connectivity are the only solutions for high performance music events.

\textbf{The network-centric perspective}: Following the network-oriented approach, Software Defined Networking (SDN) is widely used for high QoS services. SDN approach trends to be one of the most popular future Internet technologies. According to the SDN approach, the control plane logic is removed from the forwarding devices (switches and routers) that participate in the network and is incorporated into an external entity that instructs them and is called SDN controller \cite{sharma2014demonstrating}. The SDN controller has a global view of the complete network and based on this, it instructs the switches during the forwarding process. The communication between the SDN controller and the switches is feasible via OpenFlow protocol. OpenFlow allows the communication between the data plane that switches contain and the control plane that is located in the controller. Using SDN increases flexibility and adaptability in today's communication networks \cite{durner2015performance,lin2014enabling}. Also, incorporating SDN in QoS mechanisms enhances the existing application-aware resource allocation strategy. Using SDN for resource management benefits applications in cases of fast and sudden changes in network resource allocation, resulting in dynamic QoS management.

A common approach used by the network oriented researchers tries to prioritize users based on criteria such as Service Level Agreements (SLAs) or user credentials. During forwarding process, the data flows are forwarded through different queues that perform different queuing delays \cite{kumar2013user,sharma_implementing_2014,durner_performance_2015,sieber2015network,egilmez_distributed_2014, adami2015network}. Traffic destined to the high priority users is forwarded using the low delay queues and regular users occupy queues with higher queuing delay. The separation between the flows can take place according to the type of flows, the importance of the transmitted data or the fees paid to the service providers for better quality services defined by corresponding SLAs.

Another common approach used in QoS-aware networks exploits Type of Service or general traffic classification strategy. Researchers filter the traffic due to discrete features and apply different routing policies \cite{kumar_user_2013,diorio_platform_2015}. In \cite{sharma_demonstrating_2014}, traffic is grouped in two categories: business and best-effort traffic. The architecture checks the TOS field in packet header and if it is enabled, the traffic is forwarded using low delay queues. Otherwise, queues with higher delay are chosen. In \cite{hilmi_e._openqos}, the incoming traffic is grouped into data flows and multimedia flows. The type of flows is dynamically assigned on QoS guaranteed paths and data flows remain on traditional shortest paths. Criteria for traffic classification are headers used in MPLS, TOS field in IPV4, Traffic class field in IPV6, the source IP in case of a well-known multimedia server or matching pair of Transport port numbers.

Apart from traffic classification, another important feature in QoS-aware Software Defined Networking is the methodology used to measure the quality for  each path of the network \cite{alvizu2014can}. For instance, in \cite{nam_towards_2014} the proposed method initially selects the best delivery nodes. The SDN controller receives measurements from these nodes about delay and jitter. Then, the controller installs flow-rules for path between source IP and best delivery node using MPLS approach. Finally, the end node evaluates the QoE of the service and informs the SDN controller.

In \cite{jarschel2013sdn}, the SDN controller collects application state information periodically using a polling approach and due to the collected feedback, it operates network resource management in order to optimize the user experience. Moreover, another interesting approach is introduced in \cite{liotou_sdn_2015}, where they try to map QoS metrics to QoE levels. For this reason, there is a QoE-server that requests periodically QoS metrics from Mobile Network Operators (MNO) for QoE monitoring. These requests are translated into flow rules for the MNOs. The MNO asks the appropriate nodes and informs the controller to modify the policies for improved QoS. Service collaboration with the Network Providers is also proposed in \cite{bhattacharya_sdn_2013} and \cite{kumar_user_2013}, where negotiation between the controller and the Internet Service Providers takes place for QoS optimization. Feedback is collected in a distributed way and information transmission to the SDN controller is also proposed in \cite{bertaux2014dds} and in \cite{bueno2013opennaas}. QoS-related information is also collected from the controller in Real-Time Online Interactive approaches introduced in \cite{gorlatch2014improving}, where the  boundaries of QoS metrics are examined from the controller.

All the perspectives presented in section \ref{sec:related_work} study the NMP technology exclusively either from the audio/signal processing or the network perspective, ignoring each other. In our approach, we try to approach the NMP performance in terms of the end-to-end delay combining the two different perspectives and the corresponding delay types (blocking or network delay). During a live NMP event, our architecture allows interaction between the two fundamental components that participate in a NMP, the application and the network, in order to overcome network delay increase and keep the end-to-end delay below the EPT despite the congestion problem. This is feasible by modifying the audio processing pattern that results in blocking delay decrease. Following this approach, the system can afford network delay increase offering seamless quality of service.

In the current work, we combine the two perspectives, the audio-centric and the network-centric approaches, in order to achieve below the EPT end-to-end delay in an NMP system, despite the network congestion. A first draft of this work is presented in \cite{lakiotakis2017application}. In this work we use more audio modes for the audio re-configuration process and we avoid possible drops in audio quality. This approach increases the flexibility of our system in case of network congestion and simultaneously the resistance to sudden network changes, aiming to guarantee the required Quality of Service for the NMP systems.

\textbf{State-of-the-art NMP approaches}: In recent decades, there are many research approaches for the NMP research area. Apart from the categorization due to the perspective that NMP approaches use, there are many other criteria to classify similar approaches \cite{history}. The first criterion is the purpose for the NMP applications. In many approaches, NMPs are used for music teaching purposes. For this reason video and audio packets are transmitted, in order to develop the technique of the student \cite{sawchuk2003remote,carot2008distributed,drioli2013networked,akoumianakis_musinet_2014}. In these cases, high quality video is required but the overall end-to-end delay is higher than 60 milliseconds, which prevents for real-time music performances. In order to create a virtual environment like performing in the same room, some applications adjust the volume of the transmitted audio signal, trying to emulate the direction of arrival for the audio signal like performing in the same environment for each musician \cite{el2013reactive}. Also, some applications, such as \cite{kapur2005interactive}, allow text or visual feedback based on the quality of the received signal.

Other approaches focus only on the audio content transmission. Researchers exclude the video streams and allow MIDI audio data transmission. This approach refers to electronic instruments, ignoring the human voice \cite{chafe2000simplified,caceres2010jacktrip2,gabrielli2016wireless,meier2014jamberry,saputra2012design,renwick2012sourcenode,caceres2010jacktrip}. Among these approaches, some require wired Internet connectivity, but others allow wireless audio transmission using mobile devices or sensors. In \cite{taymans2013gstreamer}, the video transmission is optional for the user or only the video content is transmitted \cite{handberg2005community}. Wireless connectivity is also provided for large-scale music performances, where users occupy their mobile devices in order to create virtual audio-scenes, using the sensor network used for tracking user's location \cite{wozniewski2008large}.

Another criterion for NMP categorization is the architecture that they use for the interaction between the different audio sources. There are two main approaches towards this direction. The first relies on peer-to-peer model. Each musician transmits his/her audio stream to all other musicians. This approach is not scalable since for a large number of participants, there are excessive bandwidth requirements for the audio transmission. A solution for this problem is decreasing the audio quality or increasing the degree of compression using audio encoders in order to reduce the bit-rate and correspondingly the required bandwidth. Nevertheless, some applications adopt the peer-to-peer model for the audio transmission among the musicians \cite{drioli2013networked,caceres2010jacktrip2,carot2008distributed,akoumianakis_musinet_2014,kapur2005interactive,el2013reactive, gabrielli2016wireless, chafe2000simplified, stais2013networked,meier2014jamberry}.

The alternative approach met in NMP architectures uses the client-server model. In more detail, a central server collects all the transmitted audio streams from the musicians. It mixes them into a single stream and this stream is sent back to all participants \cite{zimmermann2008distributed,sawchuk2003remote,akoumianakis_musinet_2014,caceres2010jacktrip1,
xiaoyuan_gu_network_centric_2005,renwick2012sourcenode,saputra2012design}. This approach increases the system scalability but it requires a server with increased computational resources. Also, the centralized approach lacks redundancy and security since we have a single point of failure architecture. The mixing process, that takes place at the server side, inserts a delay overhead in the end-to-end delay. This happens because as the audio packets are received by the server, they are transferred from the kernel to the application layer, where they are replicated and they are passed again to the kernel layer in order to follow the inverse direction towards each participant. A solution to this problem is proposed in \cite{lockwood2007netfpga}, where the computationally demanding processes, such as copying, context switching and transmission are assigned to a NetFPGA, which is equipped with computational resources and memory. Another approach introduces the Netmap server, where the application layer is capable of processing directly the audio packets in kernel layer without copying \cite{rizzo2012netmap}. The last solution is presented in \cite{kohler2000click}. In this approach, the Click router is used. The Click router has a control part at the application layer and a processing/routing part at the kernel layer. The Click router enables also packet processing in the kernel layer, avoiding additional copying process and context switching.

All NMP systems, in order to guarantee high quality audio transmission, use various encoding techniques. On the other hand, various approaches prefer transmitting uncompressed audio packets. In the real-time audio transmission, there is a trade-off between high bit-rate and latency. In more detail, the encoding methods reduce bit-rate due to the available bandwidth but this process increases the audio processing delay, which is crucial in the NMPs. Other NMP models use a codec set for improving their quality. For instance, FLAC codec is used for lossless encoding purpose in \cite{xiaoyuan_gu_network_centric_2005}. A large number of approaches use ultra-low delay encoding methods, such as CELT \cite{akoumianakis_musinet_2014,carot2008distributed,gabrielli2016wireless}, ULD \cite{carot2008distributed} or OPUS \cite{carot2008distributed,meier2014jamberry}. A large fraction of the NMP applications use mono/stereo channels and fewer support multiple channel configuration.

The majority of NMPs use UDP for multimedia streaming purposes. This happens due to the lower transmission complexity that UDP \cite{UDP} introduces, compared with TCP \cite{TCP}. UDP supports smaller packet header and also does not contain mechanisms for congestion control, in-order delivery and retransmission. For this reason, in order to cope with packet loss problem in UDP cases, many NMP approaches allow the transmission of redundant data in each packet \cite{caceres2010jacktrip} or error detection/correction methods using data interpolation \cite{sinha2003loss}. Some approaches ignore error detection/correction methods in order to avoid delay increase. Other exploit protocol information (such as timestamps and sequence numbers in RTP \cite{frederick2003rtp}) in order to cope with packet-loss or out-of-order delivery problem.
\section{Conclusions and Future Work}
\label{sec:conclusions}

In the current work we introduced and demonstrated a method for close collaboration between the application and the network aiming to system performance improvement. This model can be applied in all QoS-aware applications that should provide services satisfying certain requirements related to the end-to-end delay, jitter, packet loss and error rate. We selected Networked Music Performance (NMP) systems to apply this concept, because this type of application belongs to ultra-low delay sensitive applications that have strict delay constraints at the level of milliseconds.

Our method manages to cope with a significant fraction of the network delay increase by modifying the audio configuration parameter set as we described in section \ref{sec:evaluation} and generally benefits system's performance by providing better QoS to the  users. The gain degree depends on the network conditions and the available modes that are offered by the application for switching in case of excessive network delay increase. The whole architecture is based on the recent trend in computer network area called Software Defined Networking.

Through the interaction between the application and the network, we offer better network utilization since optimal path routes are used for data delivery, avoiding traffic congestion and link failure problems. Additionally, the application can use network statistics to inform users about bad network conditions in case that it cannot cope with the problem. Generally, since for most of the recent applications Internet connectivity is indispensable for data exchange, this collaboration concept can benefit both sides. This means that the applications would be more flexible in sudden network changes without interfere users and computer networks will be used for satisfying certain application requirements described in Service Level Agreements. Moreover, Internet Service Providers can adjust their network infrastructure easily to new demands due to technological breakthroughs in coming years. In this case, the encoder and the decoder would also interact with the network and modify encoding/decoding method by changing the number of bits used for audio representation in case of traffic congestion, allowing bit-rate adoption to network condition. Additionally, apart from the audio services, we can support also video teleconferences that require instant interaction between participants located in different places. In this case, the SDN controller should assign also paths for video packets and in case of traffic congestion, it reroutes them to new paths or modify multimedia processing pattern. Moreover, a similar application type that can adopt our concept is online gaming, that requires ultra-delay interaction between online players.

\textbf{Future work}: Close collaboration between the application and the network is effective for ultra-low delay sensitive cases such as the NMP systems. The proposed Software-Defined NMP solution can be also applied in the real Internet, considering the domains (Autonomous Systems) as ``big switches'' connected with physical links or classic overlay tunneling mechanisms~\cite{kotronis2014control,Liaskos2,kotronis2016stitching}. Apart from this type of service, there are various daily life applications that can support this concept. A modified version of our system could incorporate ultra-low delay encoding/decoding methods (e.g. Opus \cite{valin2016high}) for modifying the audio bit-rate based on the network conditions.

\section*{Acknowledgement}
The work was funded by the European Research Council (ERC), project 'NetVolution', grant agreement No 338402. Also, the experimental setup was supported by GRNET S.A for providing the required computational resources for the network measurements.


\end{document}